\documentclass[11pt]{article}
\def\met{\mbox{${\hbox{$E$\kern-0.6em\lower-.1ex\hbox{/}}}_T$}\ } 

\usepackage{moriond,epsfig}
\usepackage{capt-of}
\usepackage{amsmath}
\usepackage{graphicx}

\def\Journal#1#2#3#4{{#1} {\bf #2}, #3 (#4)}

\def\PRL{\em Phys. Rev. Lett.}

\def\be{\begin{equation}}
\def\ee{\end{equation}}
\def\bea{\begin{eqnarray}}
\def\eea{\end{eqnarray}}

\begin{document}
\vspace*{4cm}
\title{Towards a Measurement of the Inclusive $W\rightarrow e\nu$ Cross Section in $pp$ Collisions at $\sqrt{s} =$ 14 TeV}

\author{ Duong NGUYEN\\
(On behalf of the CMS Collaboration)}
\address{Department of Physics, Brown University,\\
182 Hope St., Providence, RI 02912, USA}

\maketitle\abstracts{
	We present the methods for an early measurement of the inclusive $W\rightarrow e\nu$ production cross section in $pp$ collisions at $\sqrt{s} =$ 14 TeV. The methods are studied assuming 10 pb$^{-1}$ integrated luminosity of data and conditions of the Compact Muon Solenoid (CMS) detector at the early data taking period.
}
The $W$ boson inclusive production cross section measurement will be one of the first physics measurements at CMS. The theoretical predictions can be precisely calculated and the recent measurements of the two Tevatron energy frontier experiments, CDF and D\O, show agreement between theory expectations and the measured values\cite{cdfWcross}$^,$ \cite{d0Wcross}. As a consequence, the $W$ boson production cross section measurement is an important Standard Model candle for the physics commissioning at CMS. Moreover, the high transverse momentum ($p_{T}$) electron and the high transverse missing energy (\met) from the $W\rightarrow e\nu$ decay provide distinct signatures for the CMS detector calibrations, especially the Trackers, the Hadronic (HCAL), and Electromagnetic (ECAL) Calorimeter subsystem calibrations. We follow the definition of the $W\rightarrow e\nu$ cross section as:
\begin{equation}
\sigma_{W} \times BR(W\rightarrow e\nu) =  \frac{N_{W}^{pass} - N_{W}^{bkgr}}{A_{W}\times \epsilon_{W}\times\int Ldt}
\label{eqn:cross}
\end{equation}
$N_{W}^{pass}$ and $N_{W}^{bkgr}$ respectively correspond to the number of candidates selected from the data and the number of background events in the data. $A_{W}$ is the acceptance defined as the fraction of decays satisfying the geometry constrains of the detector and the pre-selection kinematic constraints. $\epsilon_{W}$ is the selection efficiency of the $W$ decays falling within the acceptance. $\int Ldt$ is the integrated luminosity. 
The data-driven methods for estimation of these quantities except the acceptance, which is calculated from Monte Carlo simulations, are discussed. The integrated luminosity measurement, which is expected to have at least 10\% accuracy from an initial Van der Meer scan of the CMS beam spot size, is out of the scope of this article.

First, the $W\rightarrow e\nu$ sample is selected by the CMS two-level trigger system, the Level-1 (L1) trigger and the High Level Trigger (HLT). The events are required to pass a trigger path which has a 12 GeV  L1 trigger transverse energy ($E_{T}$) threshold on an electromagnetic shower deposited at a cluster of ECAL crystals and a 15 GeV $p_{T}$ threshold on a reconstructed electron object at the HLT. In addition, the HLT electron is required to be isolated. Recent studies by CMS based on a trigger emulator\cite{trigger} have shown that the expected rate of the above trigger path is 17.1 $\pm$ 2.3 Hz at one of the start-up low luminosities (10$^{32}$ cm$^{-2}$s$^{-1}$) and the overall efficiency (L1xHLT) for $W\rightarrow e\nu$ events is 62\%. 

In order to reject the background and extract the $W\rightarrow e\nu$ events from the above sample selected by the trigger, selection criteria are applied to the reconstructed electrons and missing transverse energy. The reconstructed electrons need to be formed from an electromagnetic shower in the ECAL within the fiducial region ($|\eta|<$ 2.5 and 1.444 $<|\eta|<$ 1.560 excluded) and with $E_{T}\geq$ 20 GeV. Moreover, the electrons from the $W\rightarrow e\nu$ decays are isolated, thus we require a low track activity around the electron candidates which efficiently rejects electrons from the more frequent dijet events (QCD dijets) in $pp$ collisions. On top of the above criteria, we apply electron identification cuts based on a detailed simulation study of the electron reconstruction and identification\cite{eleRec} with some simplifications to obtain a high efficiency and preserve discrimination power at the early data. The \met is defined as the magnitude of the transverse vector sum over uncorrected energy deposits in the projective Calorimeter Towers\cite{METDef}. Since the reliable \met correction may not be available in the early data, we don't apply any correction to the \met, which has a mean value about 40 GeV for the $W\rightarrow e\nu$ decays, and set a cut at 20 GeV for the signal-background separation.

In order to measure the electron selection efficiency, we employ the so-called "Tag and Probe" method, which is successfully used by both Tevatron experiments and described in detail in another CMS note\cite{tagProbe}. One tag electron is required to satisfy the tight electron identification criteria, thus it is considered as a good electron candidate. The other electron, called the probe electron, is used to estimate the efficiency of passing the considered cut. The total efficiency ($\epsilon_{total}$) is factorized according to the subsequent reconstruction and selection steps of identifying an electron which are the triggering ($\epsilon_{trigger}$), the preselection ($\epsilon_{preselection}$), the isolation ($\epsilon_{isolation}$), and the electron identification ($\epsilon_{elID}$):
\begin{equation}
\epsilon_{total} = \epsilon_{trigger} \times \epsilon_{preselection} \times \epsilon_{isolation} \times \epsilon_{elID}
\label{eqn:efficiency}
\end{equation}
In a detailed CMS study\cite{tagProbe}, $\epsilon_{trigger}$, $\epsilon_{preselection}$, $\epsilon_{isolation}$ and $\epsilon_{elID}$ have the values of 0.768 $\pm$ 0.005, 0.909 $\pm$ 0.003, 0.936 $\pm$ 0.003 and 0.997 $\pm$ 0.001, respectively. The efficiency of the \met cut can be estimated from the \met model of $W \rightarrow e \nu$ events which is described below.

Another important investigation is the background estimation. The largest contribution to the background comes from the QCD hadronic dijet events, where one jet results in an electron and the other jet is mismeasured, creating missing transverse energy. There are also electroweak (EWK) backgrounds, which consist most of the $Z/\gamma^{*}\rightarrow e^{+}e^{-}$ events with one electron misidentified (3\% of signal), and the electron tau decays from $W$ and $Z$ bosons (2\% of signal). The other processes ($W\gamma, WW, WZ, ZZ, tW$) have been found to be negligible. Compared to the QCD hadronic dijet background (QCD background), the EWK backgrounds are small and the cross sections of their processes can be reliably computed, thus they can be estimated with sufficient precision from simulation.
\begin{figure}[htbp]
\begin{minipage}[tb]{0.45\linewidth}
\begin{center}
\resizebox{7cm}{!}{\includegraphics{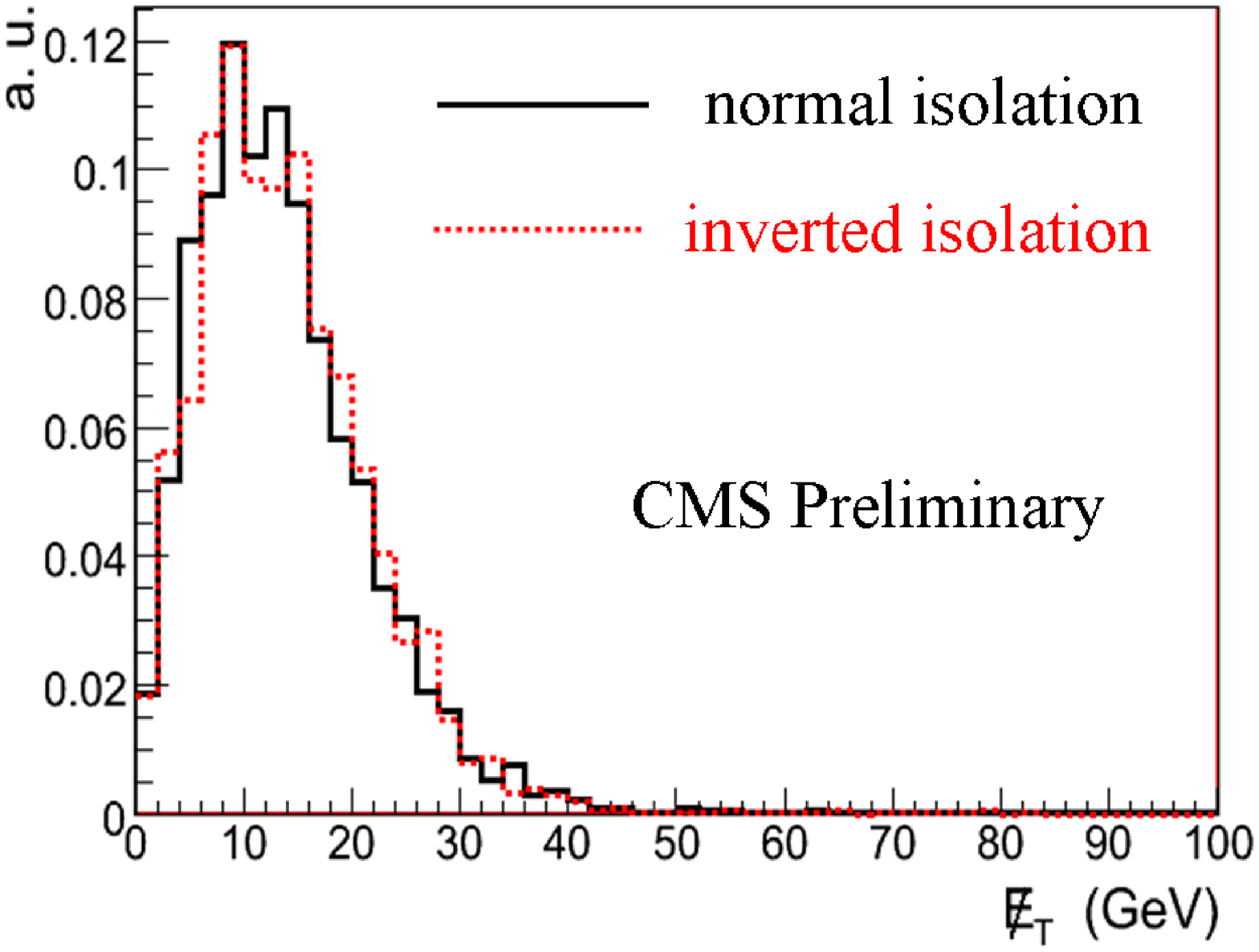}}
\caption{\met distributions of QCD dijet events which pass the electron selection with normal or inverted isolation requirements}
\label{fig:METiso}
\end{center}
\end{minipage}
\hspace{1cm}
\begin{minipage}[tb]{0.45\linewidth}
\begin{center}
\resizebox{7cm}{!}{\includegraphics{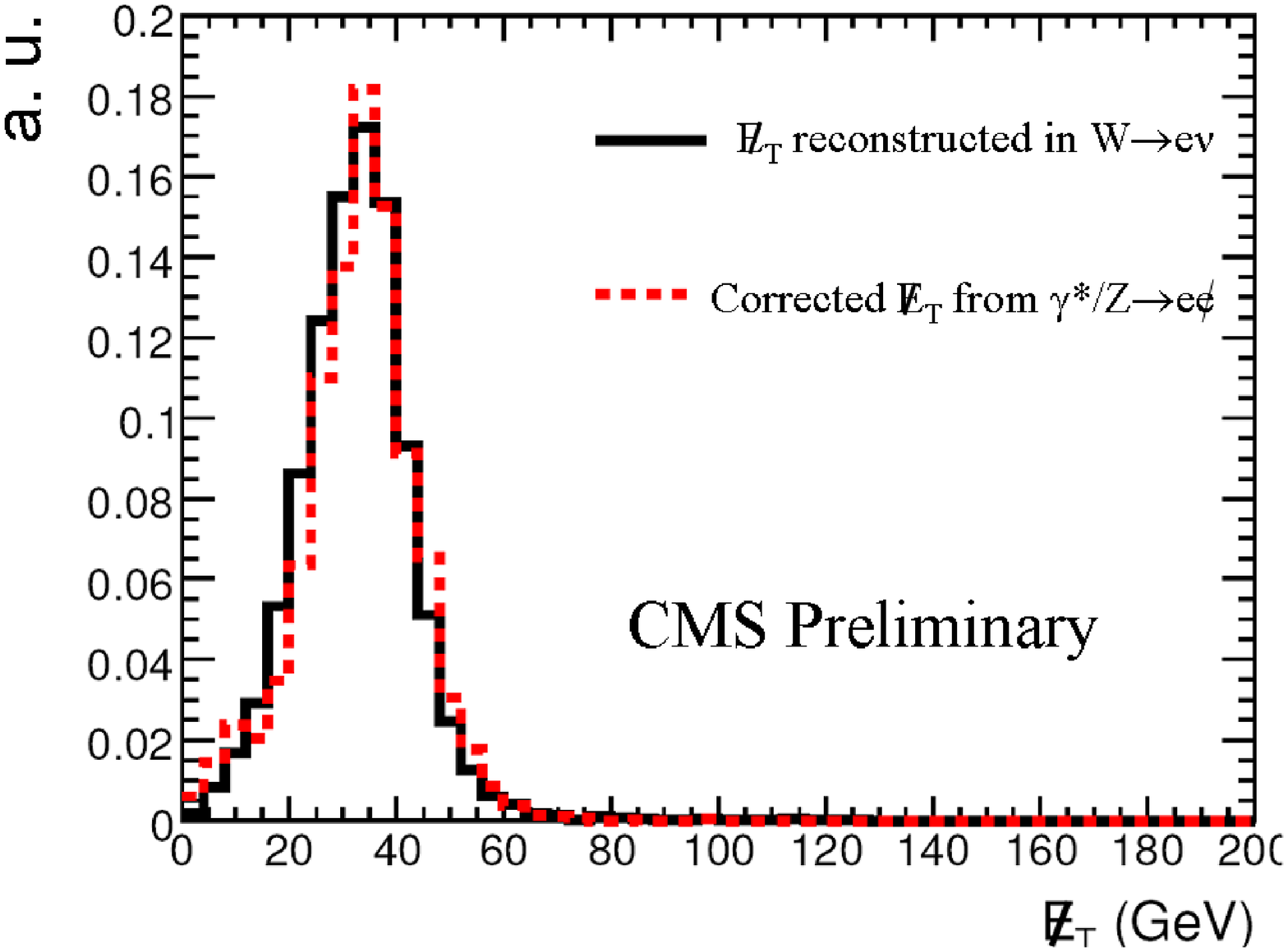}}
\caption{\met distributions reconstructed from $W\rightarrow e\nu$ events or estimated from $Z/\gamma^{*}\rightarrow e^{+}e^{-}$ events}
\label{fig:METersatz}
\end{center}
\end{minipage}
\end{figure}

Meanwhile, the QCD background arises from the strong interaction processes with large theoretical uncertainty in the cross sections. Therefore, it needs to be estimated by data-driven approaches. We use the so-called matrix method to fulfill that requirement. This method is based on the observation that the \met distribution of QCD dijet events passing the electron selection is fairly independent of the electron isolation criterion as shown in Figure ~\ref{fig:METiso}. A control sample is derived by inverting the electron isolation cut of $W\rightarrow e\nu$ selection. As a consequence, that sample consists most of the QCD background events which in many cases fail the electron isolation cut. Beside the QCD background events, the control sample contains negligible contamination from the electron-isolated $W\rightarrow e\nu$ signal events and the EWK background events.

Utilizing the above fact, from the control sample, one can derive a \met model for the QCD background events which mix with the $W\rightarrow e\nu$ signal events in the data selected by using the electron selections with a normal isolation cut. A factor, $f_{QCD}$, which is the ratio of the QCD events above and below the \met cut (20 GeV), is calculated from the \met model. Its value is $f_{QCD} =$ 0.2413 $\pm$ 0.0019 if the $W$ and other EWK background in the control sample are properly subtracted. There is an increase of 7.8\% if they are not subtracted from the control sample. Moreover, we measure a factor, $f_{Z}$, which is the same ratio as $f_{QCD}$, but estimated from a $W\rightarrow e\nu$ \met distribution model. This model is derived from the $Z/\gamma^{*}\rightarrow e^{+}e^{-}$ events by excluding calorimeter towers within a 0.1-radius cone in the $\eta-\phi$ plane around an electron. Figure ~\ref{fig:METersatz} shows that the model and true $W\rightarrow e\nu$ \met distributions agree pretty well, thus $f_{Z}$ has a value, $f_{Z} =$ 8.7 $\pm$ 0.5, which is close to the true value, $f_{W} =$ 8.13, of the $W \met$ distribution. The number of background and signal events above the \met cut in the data ($N_{>20}^{QCD}$ and $N_{>20}^{W}$, respectively) are the solutions of two equations:
\begin{equation}
\begin{split}
N_{>20}^{QCD} = f_{QCD}N_{<20}^{QCD}=f_{QCD}(N_{<20}-N_{<20}^{EWK}-\frac{1}{f_{Z}}N_{>20}^{W})\\
N_{>20}^{W}=N_{>20}-N_{>20}^{EWK}-N_{>20}^{QCD}
\end{split}
\label{eqn:bkgr1}
\end{equation}
$N_{<20}$ and $N_{>20}$ respectively correspond to the total number of events below and above the \met cut observed in the data. Their values are $N_{<20}=$ 136147 and $N_{>20}=$ 94386. The EWK background, estimated from simulation, contributes 6851 events to $N_{<20}$ ($N_{<20}^{EWK}$) and 3907 events to $N_{>20}$ ($N_{>20}^{EWK}$). Dividing $N_{>20}^{W}$ by the efficiency of the \met cut, $\epsilon(\met)=f_{Z}/(1+f_{Z})$, one can get the total number of $W$ events in the data:
\begin{equation}
N_{yield}^{W}=\frac{1+f_{Z}}{f_{Z}}N_{>20}^{W}=\frac{1+f_{Z}}{f_{Z}-f_{QCD}}(N_{>20}-N_{>20}^{EWK}-f_{QCD}(N_{<20}-N_{<20}^{EWK}))
\label{eqn:bkgr2}
\end{equation}

Applying the formula results in 67954 $\pm$ 674 background-subtracted $W$ events which is comparable with 67369 true $W$ events in the signal/background cocktail. The \met distribution of the $W\rightarrow e\nu$ signal and some sources of the backgrounds events in the data selected by the $W\rightarrow e\nu$ electron selection is shown in Figure \ref{fig:finalMET}. As can be seen, the most important background is contributed by the QCD light-flavor dijet and the heavy flavor $b\bar{b}$ events. These backgrounds are comparable to the signal, thus they need to be carefully controlled to reduce the systematic uncertainty.
\begin{table}[ht]
\begin{minipage}[tb]{0.4\textwidth}
\begin{center}
\resizebox{8.5cm}{!}{\includegraphics{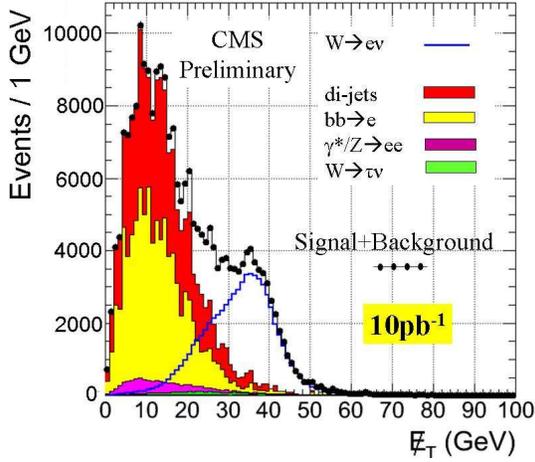}}
\captionof{figure}{\met distributions of the $W\rightarrow e\nu$ signal events and the most important backgrounds after electron selection}
\label{fig:finalMET}
\end{center}
\end{minipage}
~\hfill~
\begin{minipage}[bt]{0.5\textwidth}
\begin{center}
\begin{tabular}{cc}
\hline
$N_{yield}^{W}$ & $67954\pm674$\\
\hline\hline
Tag \& Probe $\epsilon_{offline}$ & $84.8\pm0.4\%$\\
\hline
Tag \& Probe $\epsilon_{trigger}$ & $76.8\pm0.5\%$\\
\hline
Tag \& Probe $\epsilon_{total}$ & $65.1\pm0.5\%$\\
\hline\hline
Acceptance & $52.3\pm0.2\%$\\
\hline
Int. Luminosity & $10$pb$^{-1}$\\
\hline\hline
$\sigma_{W} \times BR(W\rightarrow e\nu)$ & $19.97\pm0.25$ nb\\
\hline
Cross section used & $19.78$ nb\\
\hline
\end{tabular}
\end{center}
\caption{Results for $W\rightarrow e \nu$ cross section measurement}
\label{tab:result}
\end{minipage}
\end{table}

The results of the measurement are summarized in Table \ref{tab:result}. $\epsilon_{offline}$ is the product of $\epsilon_{preselection}$, $\epsilon_{isolation}$ and $\epsilon_{elID}$. Note that Tag \& Probe $\epsilon_{total}$ is the total electron selection efficiency which is substituted for $\epsilon_{W}$ into Equation \ref{eqn:cross}. Thus $N_{yield}^{W}$ is used to calculate the cross section. The quoted uncertainties consist of only the statistical errors calculated from number of counted events. Although the systematic uncertainties are not considered in this analysis, they are known to be mainly dominated by the luminosity measurement uncertainty which is expected to be at least 10\%. As can be seen, there is an agreement within error between the measured cross section and the assumed next-to-leading-order cross section used to generate the $W\rightarrow e\nu$ sample. Detailed description of this analysis can be found at one of the references\cite{wNote}.


\section*{References}
\begin{thebibliography}{99}
\bibitem{cdfWcross} D. Acosta et al. (CDF Collaboration), \Journal{\PRL}{94}{091803}{2005}
\bibitem{d0Wcross} D\O\ Collaboration, {\em ``Measurement of the Cross Section for $W$ and $Z$ Production to Electron Final States with the D\O\ Detector at $\sqrt{s} =$ 1.96 TeV''}, D\O\ Note {\bf 4403-CONF}, (2004)
\bibitem{trigger} CMS Collaboration, {\em ``CMS High Level Trigger''}, CMS Note {\bf 2007/009}, (2007)
\bibitem{eleRec} S. Baffioni et al. (CMS Collaboration), {\em ``Electron reconstruction in CMS''}, CMS Note {\bf 2006/040}, (2006)
\bibitem{METDef} S. Esen et al. (CMS Collaboration), {\em ``\met Performance in CMS''}, CMS AN {\bf 2007/041}, (2007)
\bibitem{tagProbe} G. Daskalakis et al. (CMS Collaboration), {\em ``Measuring Electron Efficiencies at CMS with Early Data''}, CMS AN {\bf 2007/019}, (2007)
\bibitem{wNote} CMS Collaboration, {\em ``Towards a Measurement of the Inclusive $W\rightarrow e\nu$ and $\gamma^{*}/Z\rightarrow e^{+}e^{-}$ Cross Sections in $pp$ Collisions at $\sqrt{s} =$ 14 TeV''}, CMS AN {\bf 2007/026}, (2007)
\end{thebibliography}
\end{document}